\documentclass[prl,twocolumn,preprintnumbers]{revtex4}
\usepackage{amssymb}
\usepackage{amsmath}
\usepackage{graphicx}
\usepackage{makecell}
\usepackage{multirow}
\usepackage{float}
\usepackage[caption = false]{subfig}
\begin{document}

\title{Non-Payoff Monotonic Dynamics in an Evolutionary Game of  Courtship}

\author{A. Chacoma}
\author{M. N. Kuperman}
\author{D. Zanette }

\affiliation{Centro At\'omico Bariloche (CNEA) and Instituto Balseiro}
\affiliation{Consejo Nacional de Investigaciones Cient\'{\i}ficas y T\'ecnicas, \\
(8400) Bariloche, R\'{\i}o Negro, Argentina.}

\begin{abstract} We propose an evolutionary coordination game to formalize
a simplified model of the evolution of strategies during human
courtship. The dynamics, derived from the consideration of
experimental observations on human social behavior driven by
self-esteem, turns out  to be non-payoff monotonic. This property
gives rise to nontrivial evolution in the players' strategies, which
we study  both numerically and analytically.
\end{abstract}

\maketitle

\section{Introduction}
\label{intro}
Nature offers  a plethora of instances showing that relations
between sexes are not always smooth. Extreme examples are the
praying  mantis female that cannibalizes its mate; the male bedbug,
that slashes the female's abdomen with a sharp sting; and the
toxin-loaded scorpion, that inoculates a special venom to subdue the
female of the species. Aside from these drastic cases, most species
show conflicting behavioral patterns between males and females when
involved in a mating relation.  In  general, male behavior has
evolved to crave scoring at any costs. On the contrary, the female
is inclined to ensure  viable offspring. This situation leads to a
conflict between sexes concerning their respective shares in
parental investment.  Offspring raising requires a considerable
amount of time and energy, that each parent might attempt to
minimize  at the expense  of the other.

Humans, for example, require intense cooperation to raise  children
to adulthood \cite{hrdy}. They  have evolved as cooperative
breeders, and  the presence of a helper is often essential to be
able to raise offspring.  But the conflicts that arise from breeding
are not unique to  humans. Whenever fertilization is internal
\cite{sigm}, females risk being deserted even before  giving birth
to their offspring.  Furthermore, as females produce relatively few,
large gametes, whereas males produce many small gametes, the former
are much more committed to breeding while the later are  in a better
position to desert. Males can invest time and energy not used in
breeding into increasing their offspring with numerous mates. Male
strategy is thus a fast and short engagement with the female.
Females need adopting a counter-strategy to avoid  desertion, and
they insist upon a long engagement period before copulation. The
complexity of the situation is interesting enough to attempt
mathematical modeling.  Game theory provides  a suitable frame to
discuss the so-called Battle of the Sexes \cite{luce} as a strategic
game. Formulating a $2\times 2$ game (two players, with two
strategies each) where the male strategies are philandering and
faithful, whereas the female strategies are  coy and fast, it was
shown that in a population of coy females, males must be faithful
\cite{sigm}. Among faithful males, however, it would not pay a
female to be coy. This is an example of a coordination game, a class
of games that we discuss in Section \ref{s2}.

In most vertebrate species,  the engagement of a male and a female
is preceded by courtship. The courtship process in humans has been
analyzed by psychologist in several studies \cite{freu}. In
particular, in the first stage of courtship  both genders have equal
responsibilities in the successful evolution of the process. During
their initial approach, the man and the woman make   decisions based
on parameters  defined by their social conception and personality,
which  affect the subsequent course of the interaction. It was
experimentally observed that, in general, women look for  obtaining
a  stable relationship, unlike men who try to seduce the female
seeking  a sexual partner \cite{hen}. These dominant behaviors might
have been socially shaped, and seem to be rooted on evolutionary
paths.

In this paper, we analyze a simplified version of this problem. We
consider two  interacting populations:  males and females. While
individuals of both sexes seek to engage in a relationship with
individuals of the opposite sex, their expectations are different.
Males look for a  direct approach to females, to get involved as
soon as possible in a relation. Females, on the other  hand, look
for a more cautious approach, let seduction flourish, test their
potential partners. We consider that both individuals will profit
from establishing a relationship, that will prosper only if both
partners match their respective expectations or strategies. A male
will abandon his direct intent if the female is not immediately
responsive, and a cautious female will generally not accept the
approach of a man with no further intentions than sexual
intercourse. After revisiting the concept of coordination game in
Section \ref{s2}, a mathematical formalization of the problem will
be presented in Section \ref{s3}.

The choice of a courtship strategy is expected to require highly
adaptive decision-making abilities \cite{lee}. Among the constraints
that affect this choice, we focus on self-esteem, which can be
defined as  the set of thoughts, perceptions, feelings about
ourselves, built up out of  the collected experience we grasp in
every  social interaction along our lives  \cite{jan}.  This
self-evaluation directly affects the way how we relate to others and
how we behave in social environments. People with different levels
of self-confidence, respond differently when facing the same social
challenges  \cite{clar,anto}. Self-esteem, moreover, increases when
an individual enlarges his/her record of successful interactions. In
the case of courtship, this fact promotes the direct approach
strategy among men and the opposite strategy among  women, who
become more selective when choosing their partners. In other words,
the more successful the  players are the more inclined they are to
play their favorite strategy. In order to represent these aspects we
introduce evolutionary game rules where the probability that an
individual adopts either strategy depends on self-esteem. In turn,
self-esteem is a function  of  the player's accumulated experience
or score. Additionally, as a way of including social influence
effects, we also consider the possibility of strategy imitation
between males.

\section{Coordination games}
\label{s2}

A coordination game represents a situation in which players can make
their strategies agree with each other in order to get the highest
payoff. In a $2\times 2$ game, suppose that  player A  can adopt
strategies A1 and A2, and player B  can adopt strategies B1 and B2.
Coordination is possible when the payoffs, shown in Table
\ref{Table1}, satisfy $A > B$, $\Delta > \Gamma$, and  $\alpha >
\gamma$, $\delta > \beta$. In this case, the strategy combinations
\{A1, B1\} and \{A2, B2\} are pure Nash equilibria  \cite{sigm}.

\begin{table}[h]
\begin{center}
\begin{tabular}{|c|c|c|}\hline
\diaghead{\theadfont ColumnmnHead } {player A}{player B}&
B1&B2\\
\hline
&&\\
A1& \ \ $A,\alpha$ \ \ & \ \ $\Gamma,\gamma$\ \  \\
&&\\
\hline
&&\\
A2& $B,\beta$ & $\Delta,\delta$ \\
&&\\
\hline
\end{tabular}
\end{center}
\caption{Payoff matrix of a general $2\times 2$ game. In each cell,
the two figures give, respectively, the payoff of player A and
player B for each pair of strategies.} \label{Table1}
\end{table}

Apart from these two pure Nash equilibria, and depending of the
relative values of the payoffs, the coordination game may also have
mixed Nash equilibria. It can be shown that the mixed strategy where
player A chooses A1 with probability
$$\mu =(\delta-\beta)/(\alpha+\delta-\beta-\gamma)$$
and A2 with probability $1-\mu$, and player B chooses B1 with probability
$$\nu = (\Delta-\Gamma)/(A+\Delta-B-\Gamma)$$
and B2 with probability $1-\nu$, is also a Nash equilibrium. Unlike
the pure Nash equilibria, however, the mixed equilibrium is not an
evolutionarily stable strategy \cite{may}. This can be proven by
considering the replicator equations of the game  \cite{sigm}, which
state that the probabilities $\mu$ and $\nu$ evolve according to
\begin{equation} \label{repl}
\begin{array}{ll}
\dot \mu = \mu \left( \phi_{\rm A1} - \bar \phi_{\rm A} \right),
&\mbox{with  }  \bar \phi_{\rm A} = \mu \phi_{\rm A1} + (1-\mu)
\phi_{\rm A2} ,
 \\ \\
\dot \nu = \nu \left( \phi_{\rm B1} - \bar \phi_{\rm B} \right),
&\mbox{with  } \bar \phi_{\rm B} = \nu \phi_{\rm B1} + (1-\nu)
\phi_{\rm B2} ,
\end{array}
\end{equation}
where $\phi_{\rm A1}$,   $\phi_{\rm A2}$,    $\phi_{\rm B1}$,   and
$\phi_{\rm B2}$,   are the average payoffs of individual strategies,
namely
\begin{equation}
\begin{array}{ll}
\phi_{\rm A1} = \nu A + (1-\nu) \Gamma  , &  \quad
\phi_{\rm A2} = \nu B + (1-\nu) \Delta   \\ \\
\phi_{\rm B1} = \mu \alpha + (1-\mu) \beta  ,  &
\quad  \phi_{\rm B2} = \mu \gamma + (1-\mu) \delta  .
\end{array}
\end{equation}
The Nash equilibria of the game  are also fixed points of the
replicator equations,  while stable fixed points of the equations
are Nash equilibria of the game. Equations (\ref{repl}) has five
fixed points, four of them corresponding to all the possible
combinations of pure strategies, and the  fifth corresponding to the
mixed strategy. Among them, only the pure  Nash equilibria \{A1,
B1\} and \{A2, B2\}  are stable.

In the following, we consider a subclass of coordination games,
generally known as the  Battle of the Sexes \cite{luce}. In the
traditional formulation of this  game, two players --Alice and Bob,
say-- try to decide how to  spend their holidays together. Alice
prefers going to the beach, while Bob is interested in mountain
hiking. Despite this disagreement, however, both would prefer to go
to the same place rather than to different ones.  The Battle of
Sexes is defined by the payoff relations $A=\delta>\alpha=\Delta$.

\section{The game of courtship}
\label{s3}

As a special case of the Battle of Sexes, we study a game with two
players, male and female, each of them with two available
strategies, direct (D) and indirect (I). In connection with the
courtship process described in the Introduction, the two strategies
are respectively related to more open  and more reserved attitudes.
The payoff matrix is shown in Table \ref{Table2}. While the
coordinated strategies \{D,D\} and \{I,I\}, which we associate with
mutually successful courtship, are more convenient for the two
players than the other two combinations, it is clear that the male
prefers D and the female prefers I.

\begin{table}[h]
\begin{center}
\begin{tabular}{|c|c|c|}\hline
\diaghead{\theadfont ColumnmnHead }%
{male}{female}&
D&I\\
\hline
&&\\
D& \ \ $2,1$ \ \ & \ \ $-1,-1$\ \  \\
&&\\
\hline
&&\\
I& $-2,-2$ & $1,2$ \\
&&\\
\hline
\end{tabular}
\end{center}
\caption{Payoff matrix of the Battle of Sexes considered in the text.}
\label{Table2}
\end{table}

\begin{figure}
\subfloat[]{\includegraphics[width=4.4cm, clip=true]{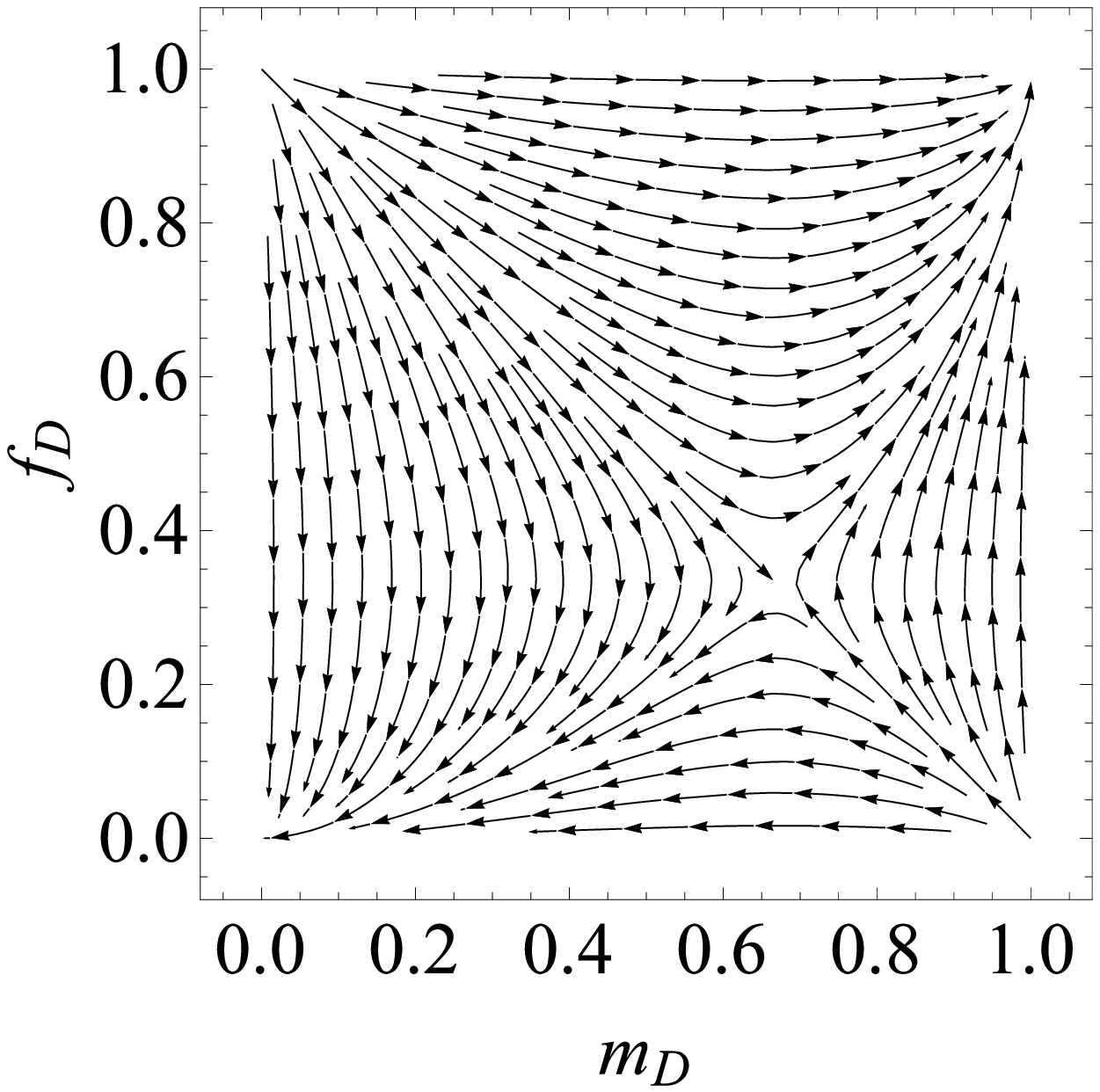}}
\subfloat[]{\includegraphics[width=4.4cm, clip=true]{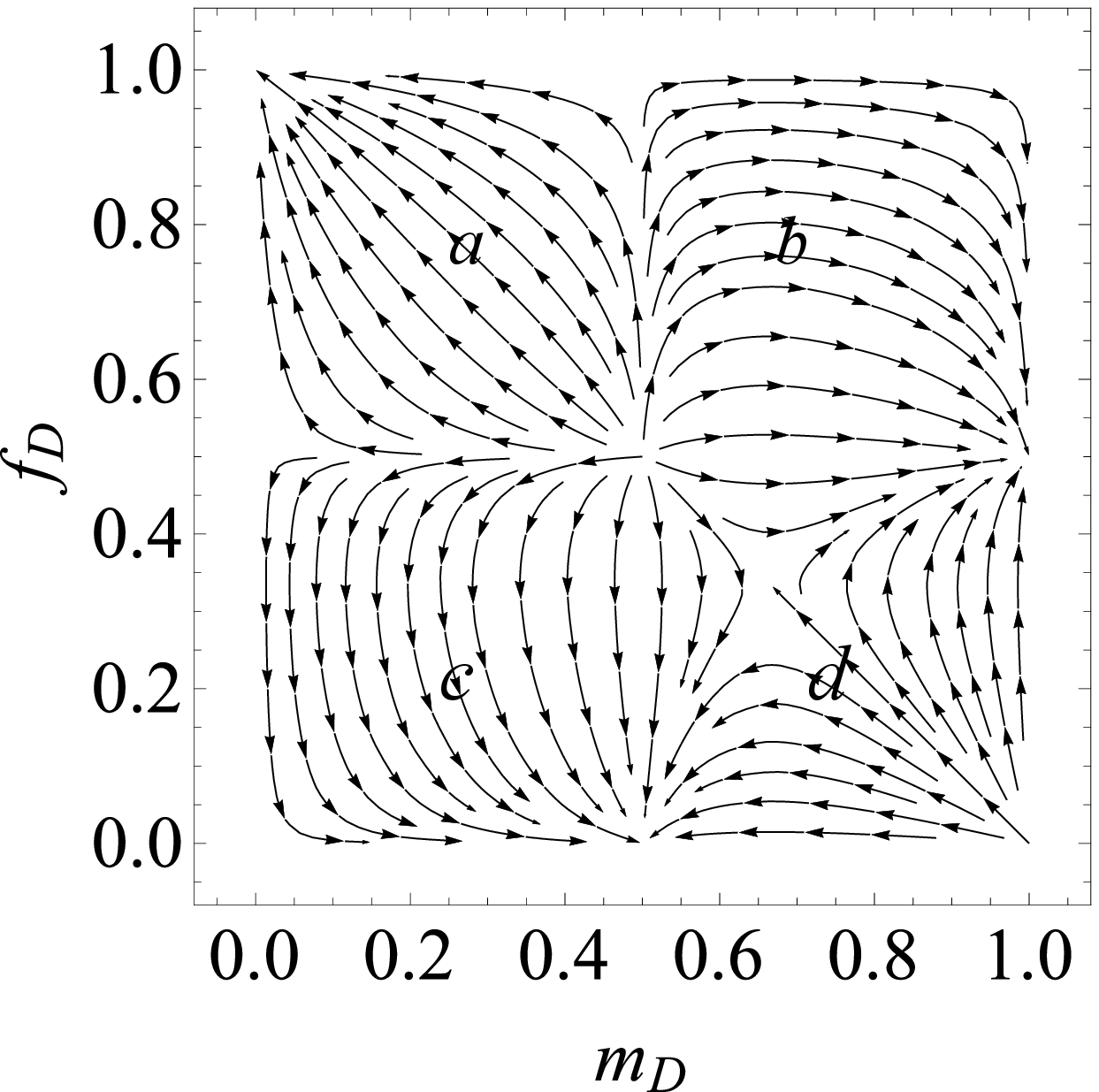}}
\caption{Flow plots for the game of courtship in the plane of direct
strategy probabilities for male ($m_{\rm D}$) and female ($f_{\rm
D}$) corresponding to (a) replicator dynamics, Eqs.~(\ref{repli2}),
and (b) self-esteem dynamics, Eqs.~(\ref{influ2}). Labels in the
right panel indicate the zones where the initial conditions for the
numerical realizations shown in Fig.~\ref{bias} were chosen.}
\label{and1}
\end{figure}

In this game, the replicator equations for the probabilities $m_{\rm
D}$ and $f_{\rm D}$ that respectively males and females play D are
\begin{equation}
\label{repli2}
\begin{array}{ll}
\dot m_{\rm D} &= 2 m_{\rm D}(1- m_{\rm D})(3 f_{\rm D}-1),  \\
\\
\dot  f_{\rm D}&= 2  f_{\rm D}  (1- f_{\rm D})(3  m_{\rm D}-2) .
\end{array}
\end{equation}
As in  Section \ref{s2}, we can verify the existence of a mixed Nash
equilibrium  with  probabilities $m_{\rm D}=2/3$ and  $f_{\rm
D}=1/3$, apart from the pure Nash equilibria $\{ 0,0\}$ and $\{
1,1\}$. In Fig.~\ref{and1}a, we show a flow plot  on the $(m_{\rm
D},f_{\rm D})$-plane corresponding to  Eqs.~(\ref{repli2}).

As stated in the Introduction, however, we are interested in a
different type of dynamics, inspired on the effects of self-esteem
during courtship. We assume that  each individual  plays the
preferred strategy (D for male and I for female) with more frequency
as his/her self-esteem grows. Moreover, we propose that  self-esteem
is determined by  the degree of success (i.e., the total payoff) in
previous matches. The probability $m_{\rm D}$ of a male playing D is
associated to confidence, and should increase as his total payoff
grows.  Meanwhile, the probability $f_{\rm D}$ of a female playing D
is related to her predisposition to open courtship, and should thus
decrease with her total payoff. Calling $\Phi_{\rm M}$ and
$\Phi_{\rm F}$ the respective total payoffs of  male and female, and
taking into account that the probabilities $m_{\rm D}$ and $f_{\rm
D}$ must be bounded to the interval $[0,1]$, we choose the sigmoidal
functions
\begin{equation}
\label{influa}
\begin{array}{ll}
m_{\rm D} &=\frac{ 1}{  2}\left[ 1+\tanh(\Phi_{\rm M}-1/2)\right], \\
\\
f_{\rm D}&=\frac{1}{ 2}\left[1-\tanh(\Phi_{\rm F}-1/2)\right].
\end{array}
\end{equation}
Variations in $m_{\rm D}$ and $f_{\rm D}$ are thus governed by the
evolution of $\Phi_{\rm M,F}$ which, in turn, is given by the payoff
in each match. Specifically, we have
\begin{equation}
\dot m_{\rm D} =\dot \Phi_{\rm M} \frac{  d m_{\rm D}}{ d\Phi_{\rm
M}} = \phi_{m_{\rm D}} \frac{  d m_{\rm D}}{ d\Phi_{\rm M}} ,
\end{equation} and
\begin{equation}
\dot f_{\rm D} =  \dot \Phi_{\rm F} \frac{df_{\rm D}}{d\Phi_{\rm F}}
=  \phi_{f_{\rm D}} \frac{df_{\rm D}}{d\Phi_{\rm F}}  ,
\end{equation}
where $\phi_{m_{\rm D}}$ and  $\phi_{f_{\rm D}}$ are the average male and
female payoffs when playing D with probabilities $m_{\rm D}$ and $f_{\rm D}$ respectively. Putting
all these expressions in terms of  $m_{\rm D}$ and $f_{\rm D}$, we find
\begin{equation}
\label{influ2}
\begin{array}{ll}
\dot m_{\rm D} &=2m_{\rm D} (1-m_{\rm D})(1-2m_{\rm D}-
3f_{\rm D}+6 m_{\rm D} f_{\rm D})  ,
\\ \\
\dot f_{\rm D} &=2 f_{\rm D} (1-f_{\rm D})(-2+3m_{\rm D}+4f_{\rm
D}-6m_{\rm D} f_{\rm D}) .
\end{array}
\end{equation}
These equations have ten fixed points, with five of them coinciding
with those of  Eqs.~(\ref{repli2}). Figure \ref{and1}b shows the
corresponding flow plot on the  $(m_{\rm D},f_{\rm D})$-plane.

Remarkably,  the Nash equilibria of the Battle of Sexes are no
longer stable under Eqs.~(\ref{influ2}). This draws an important
distinction between the replicator dynamics and the self-esteem
dynamics introduced here. The replicator dynamics is a special case
of a broad family of imitation dynamics  called {\em payoff
monotonic}  \cite{wei}. The most general form of imitation dynamics
is given by the equations
$$\dot x_ i = x_i g_i(\vec x),\quad i=1,\dots , n,$$
where $x_i$ is the probability of playing a given strategy $i$ among
a set of $n$ strategies, and the functions $g_i$ satisfy
$$\sum_{i=1}^n x_i g_i(\vec x)=0,$$
on the $n$-dimensional simplex $S_n=\{\vec x \in R^n \mid x_i>0
\wedge \sum_i x_i=1 \}$. The condition for this dynamical system to
be payoff monotonic is
$$g_i(\vec x) > g_j(\vec x) \Leftrightarrow a_i(\vec x) > a_j(\vec x)$$
for all $i,j$.
This means that the  probability of playing a given strategy
increases more rapidly if the corresponding payoff is higher, at the
expenses of the less successful ones. In a series of well-known
theorems \cite{sigm,wei}, it is proven that for payoff monotonic
equations (i)  Nash equilibria are fixed points, (ii) strict Nash
equilibria  are asymptotically stable, and (iii) fixed points that
are stable are Nash  equilibria. Since these statements are not true
for Eqs.~(\ref{influ2}), we conclude that our courtship dynamics is
non-payoff monotonic, even when it was derived from social
behavioral considerations.

\section{Numerical Results and the Effects of Imitation}

Equations (\ref{influ2}) were formulated assuming that the total
payoff of any single individual evolves according to the average
payoff at each match. Therefore, they do not necessarily capture the
effects of ``microscopic''  fluctuations. Strictly speaking,
moreover, they describe the evolution of strategy probabilities for
a single male-female couple, while we are interested at analyzing
the collective social dynamics of a population where many
individuals of the two sexes have repeated opportunities of engaging
into the courtship process. These aspects can be explored by
performing agent-based numerical simulations.

We consider different population sizes ranging from $200$ to $2000$,
with  an equal number of males and females. Initially, each male or
female strategy is defined by a probability of playing the direct
strategy D, $ m_{\rm D}$ and  $f_{\rm D}$.  Through
Eqs.~(\ref{influa}), these initial probabilities define the
respective initial (preexistent) total payoffs. Here, we  analyze
several situations with localized and uniformly  distributed initial
strategies.

\begin{figure}
\includegraphics[width=\columnwidth, clip=true]{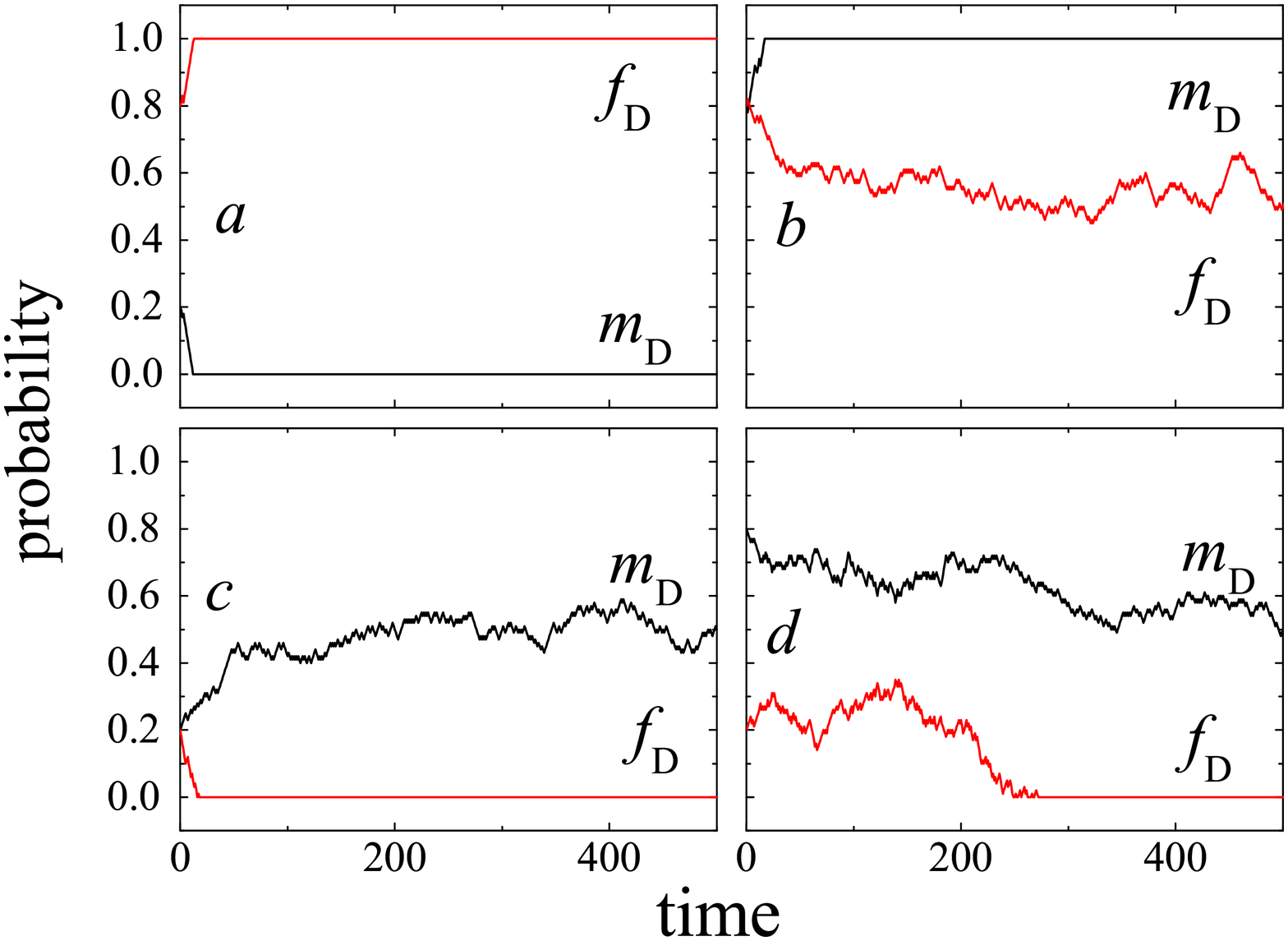}
\caption{Four numerical realizations of the game of courtship for
initial conditions in different zones of Fig.~\ref{and1}. Labels
{\it a} to {\it d} indicate the corresponding zones.}
\label{bias}
\end{figure}

First, we study the evolution of a population
with strategies concentrated in each of the four regions indicated
in Fig.~\ref{and1}b. From any of these initial conditions, the
surviving strategy profile  corresponds to one of the stable fixed
points of Eqs.~(\ref{influ2}). Starting from the conditions
indicated as $a$, the system converges to $(0,1)$; starting from
$b$, to $(1,1/2)$; and  starting from $c$ and $d$,  to $(1/2,0)$.
Figure \ref{bias}b shows typical evolutions corresponding to single
simulations with $2000$  individuals. For these localized initial
conditions, hence, numerical simulations agree with the expected
behavior derived from the equations.

\begin{figure}
{\includegraphics[width=4.4 cm, clip=true]{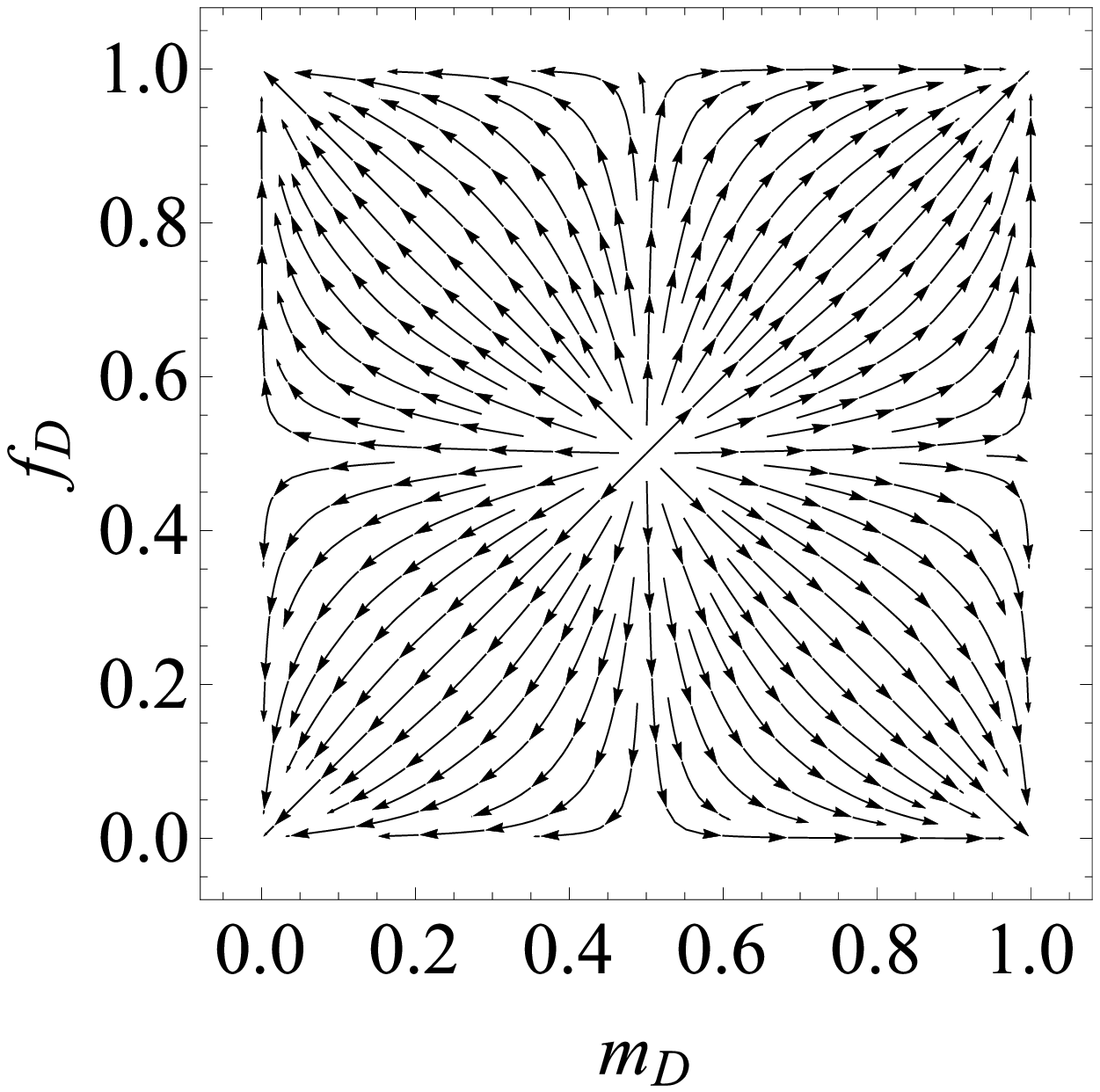}}
{\includegraphics[width=4.3 cm]{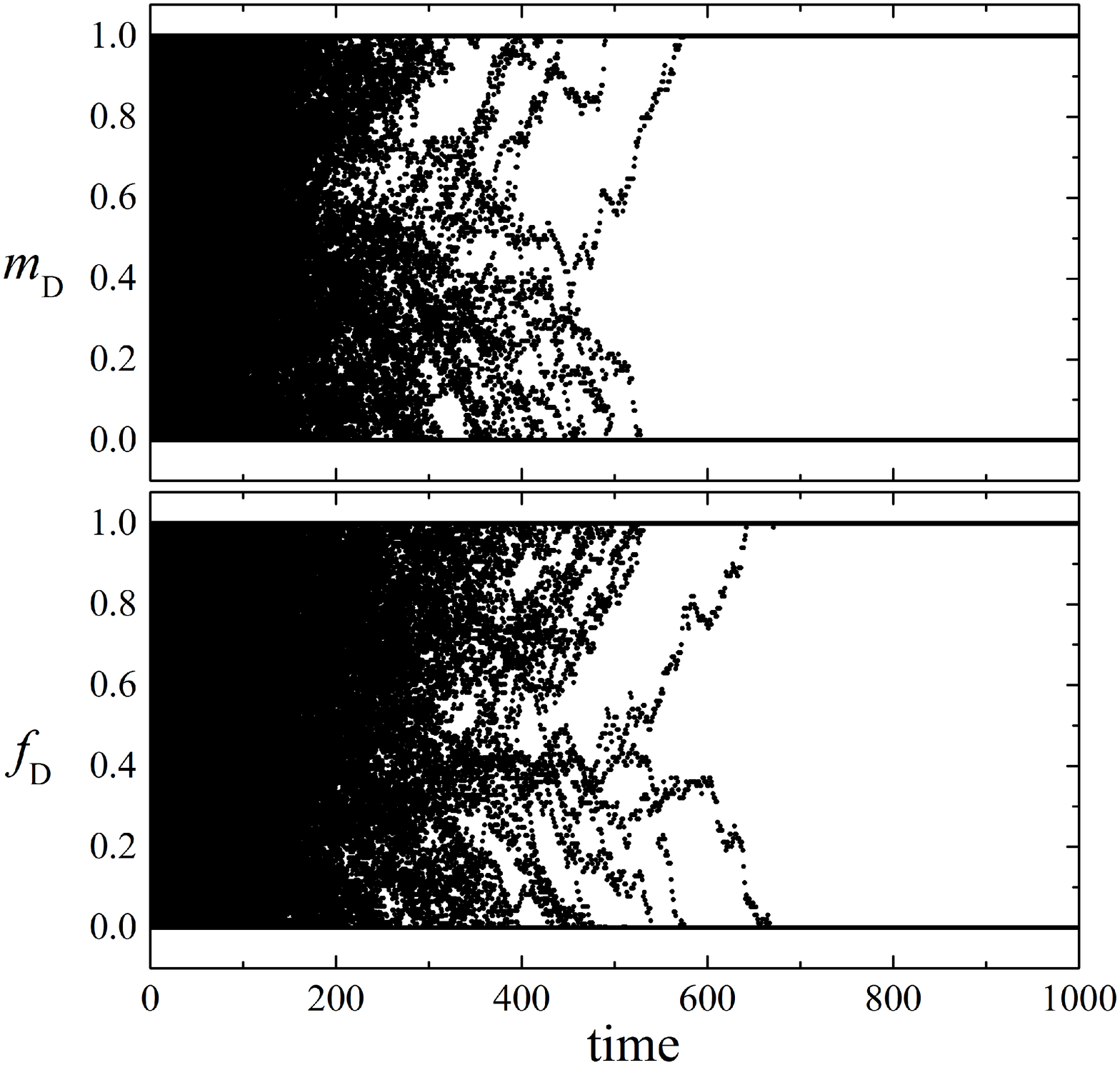}} \caption{Left panel: Flow
plot for Eqs.~(\ref{meanfi}). Right panel: Evolution of direct
strategy probabilities in a numerical simulation for a population of
$2000$ individuals, for males ($m_{\rm D}$) and females ($f_{\rm
D}$). The initial probabilities were uniformly distributed over the
interval $(0,1)$.} \label{unifor}
\end{figure}

The situation is however different when we start  from uniformly
distributed probabilities. In this case, numerical simulations show
that the male population splits into two approximately equal groups,
one of them converging to $m_{\rm D}=0$ and the other to $m_{\rm
D}=1$. The female population exhibits the same behavior with respect
to $f_{\rm D}$. This result can be understood by noting that,  due
to the symmetry of the initial condition, the system behaves at all
times as if each male and female sees the population of the opposite
sex as adopting the average strategy,  respectively, $f_{\rm D}=1/2$
and $m_{\rm D}=1/2$. In fact, replacing these values in each of
Eqs.~(\ref{influ2}), we get
\begin{equation}
\label{meanfi}
\begin{array}{ll}
\dot m_{\rm D} &=2 m_{\rm D}(1-m_{\rm D})(1/2-m_{\rm D})  ,\\
\\
\dot f_{\rm D}&=2 f_{\rm D}(1-f_{\rm D})(1/2-f_{\rm D})  .
\end{array}
\end{equation}
The corresponding flow plot, shown in Fig.~\ref{unifor}a, agrees
with the numerical observation. Figure \ref{unifor}b displays the
evolution of individual strategies in a single realization.

In order to take into account the effect of social influence, in the
form of imitation,  we introduce the probability $q$ that, at each
match, the participating male chooses its strategy by adopting the
probability of another, randomly chosen male, instead of his own.
For $q=0$, we reobtain the original model and, in fact, as $q$
remains small enough the collective behavior is the same as before.
In particular, from uniformly distributed probabilities, the
populations of both sexes split into similar groups that
asymptotically approach the extreme values. For the sake of
comparison with other instances,  we call this asymptotic behavior
case $A$.

\begin{figure}
{\includegraphics[width=\columnwidth, clip=true]{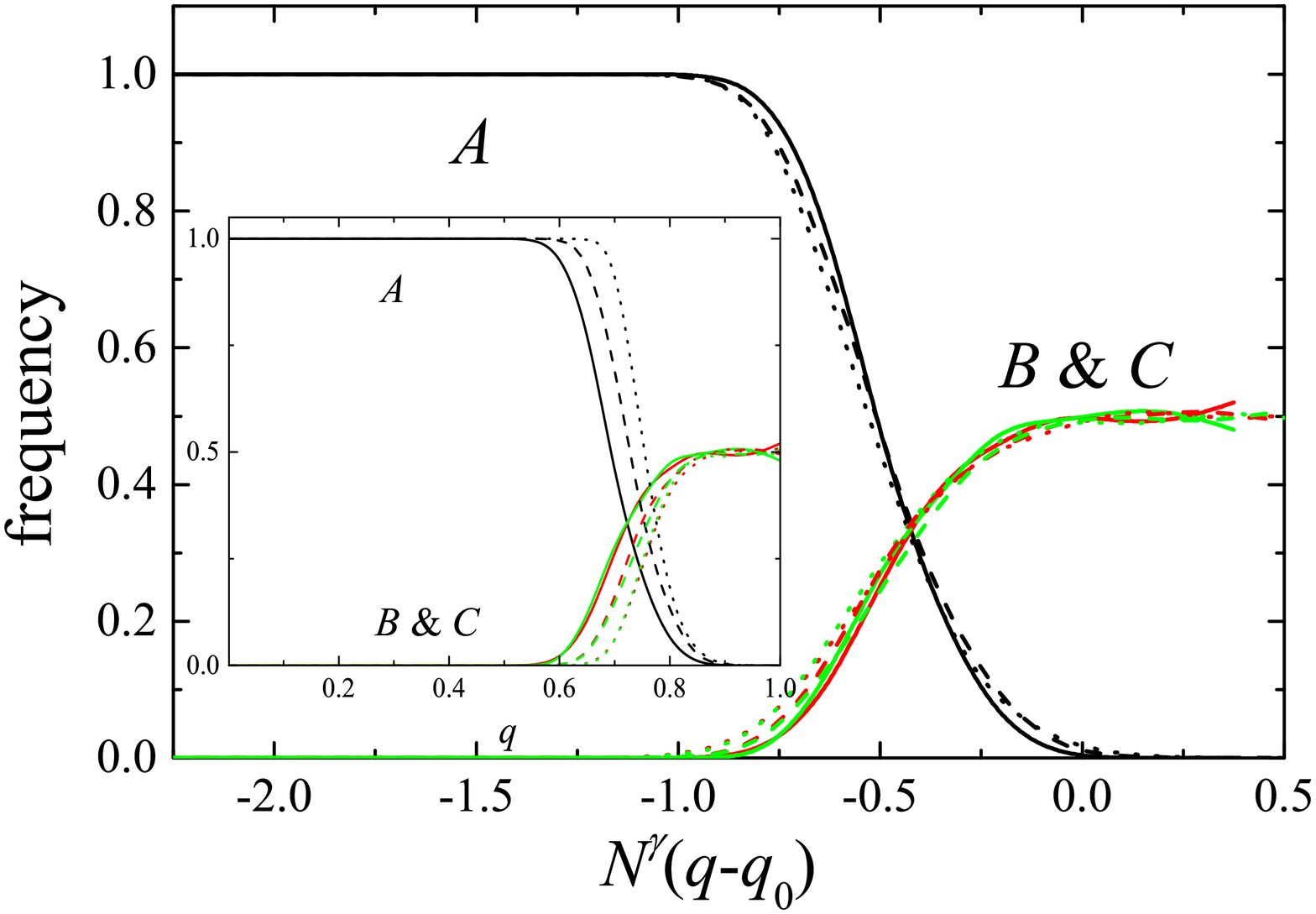}}
\caption{Frequencies of each one of the three possible asymptotic
states described in the text ($A$, $B$, and $C$) as a function of
the imitation probability $q$, out of $1000$ realizations for each
value of $q$. Different curves correspond to population sizes of
$N=200$ (full line), $600$ (dashed line) and $2000$ (dotted line)
individuals. The main plot presents a finite-size analysis collapse
of the results, with $\gamma=0.2$ and $q_0=0.87$, while the inset
shows the raw results.} \label{colap}
\end{figure}

As the imitation probability $q$ becomes larger, however, other
kinds of asymptotic behavior emerge. In some of the realizations,
all the male population converges to $m_{\rm D}=0$ while the female
population still splits into two groups that converge to extreme
probabilities  (case $B$). Alternatively, all the males converge to
$m_{\rm D}=1$ and all the females to $f_{\rm D}=1/2$ (case $C$). As
$q$ grows, the frequency of case $A$ decreases monotonically, while
the frequencies of cases $B$ and $C$, which are statistically
identical to each other, increase. Figure \ref{colap} shows the
fraction of realizations that end up in each of the three cases, as
a function of $q$. We have run $1000$ simulations for each value of
$q$. The inset shows the results for different population sizes,
while  the main plot shows a collapse of these results obtained from
standard finite-size scaling. A well-defined transition at $q_0
\approx 0.87$, with a critical exponent $\gamma = 0.2$, is apparent.

As expected, thus, sufficiently frequent imitation between males
leads their whole population to a common pure strategy. From a
uniformly distributed initial condition, the two extreme values
$m_{\rm D}=0$ and $1$ are equally probable, suggesting that the
asymptotic state is controlled by microscopic fluctuations in the
initial distribution of strategies. It is interesting that these two
asymptotic pure strategies have a mixed collective counterpart in
the female population: for $m_{\rm D}=0$, each female indistinctly
adopts either   $f_{\rm D}=0$ or $1$, while for   $m_{\rm D}=1$, all
females choose their strategies with equal probability,    $f_{\rm
D}=1/2$. This behavioral diversity, no doubt, is to be ascribed to
the non-payoff monotonic nature of the courtship game.

\section{Analytic Approach}

In the following, we introduce a mean field-like set of differential
equations for the system studied numerically in the preceding
section. Our aim is to provide an analytical approach to the results
obtained when social influence is included, and more specifically to
check the existence of a transition between two different regimes,
as shown in Fig. \ref{colap}. We have seen that, when $q=0$, the
mean values of the probabilities of playing direct for each sex are
$\overline{m}_D=\overline{f}_D= 1/2$. These values are also valid
for the case $A$, while cases $B$ and $C$, we have found
$(\overline{m}_D,\overline{f}_D) = (0,1/2)$ and $(1,1/2)$,
respectively.

We assume that the dynamical behavior of each individual can be
understood as if each player is confronting a  virtual opponent who
adopts the mean strategy of the population. Thus, each male plays
against a female adopting $f_D=\overline{f}_D$, while  each female
plays against a  male adopting $m_D=\overline{m}_D$. Moreover, the
tendency of males to be affected by the social environment is
quantified by the parameter  $q$ that defines the probability of
males of playing $\overline{m}_D$ instead of their own strategy.
With these assumptions, our mean field-like equations read
\begin{equation}
\label{qm} \dot m_D=(m_D-1) m_D ( 3\overline{f}_D-1) [1 - 2(1-q) m_D
 - 2q \overline{m}_D ]\equiv F(m_D)
\end{equation}
with fixed points at $m_D=0$, $m_D=(1-2q\overline{m}_D )/2(1-q)$,
and $m_D=1$ and
\begin{equation}
\label{qf} \dot{f_D}=(3 \overline{m}_D-2)  f_D (f_D-1 )(2
f_D-1)\equiv G(f_D)
\end{equation}
with fixed points at $f_D=0$, $f_D=1/2$ and $f_D=1$.

\begin{figure}
{\includegraphics[width=8cm, clip=true]{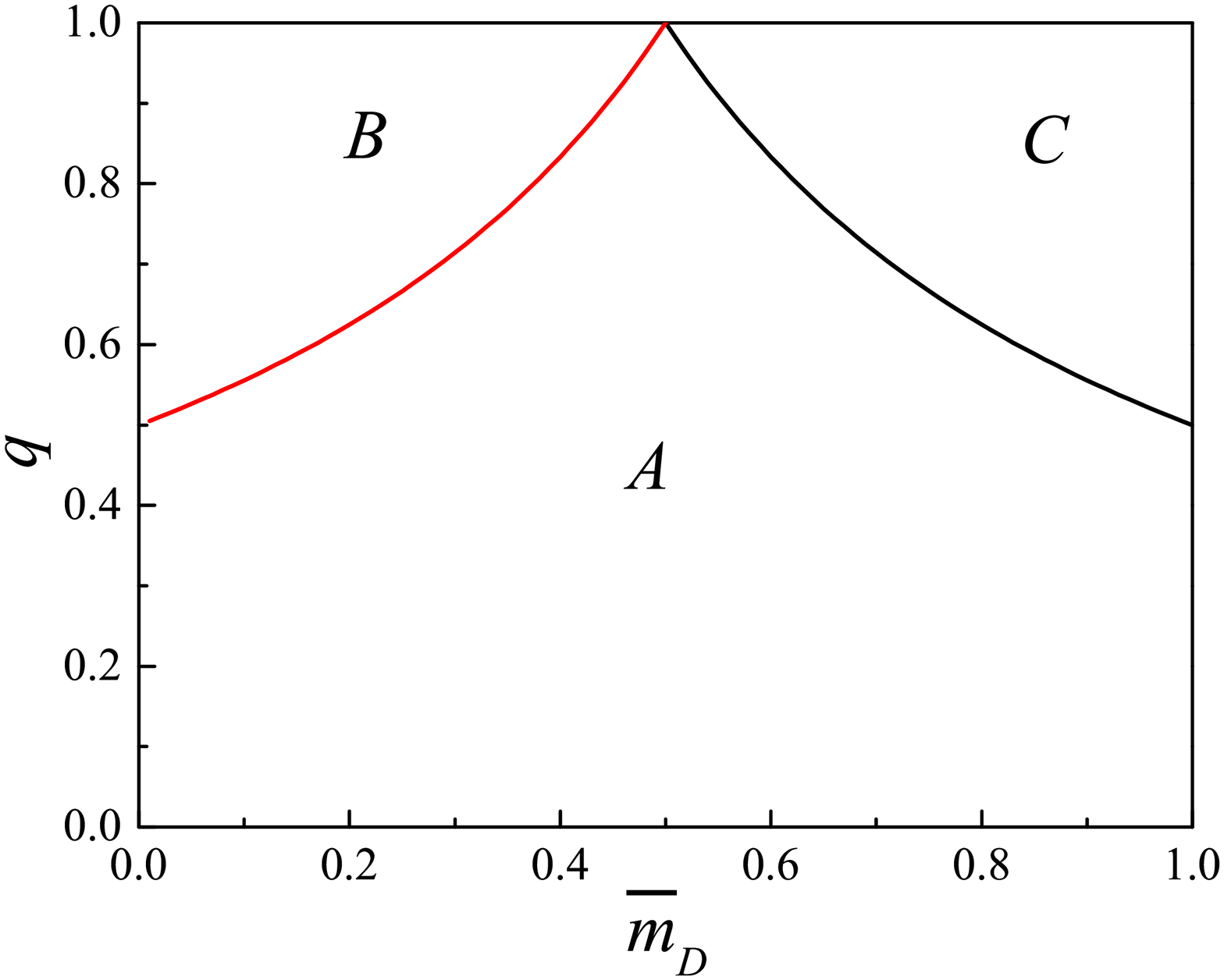}} \caption{Regions
in the plane $(\overline{m}_D,q)$ corresponding to different
stability properties of Eq. (\ref{qm}). ($A$) $m_D=0$ and $m_D=1$
are stable; ($B$)  only $m_D=0$ is stable; ($C$) only $m_D=1$ is
stable.} \label{stab}
\end{figure}

Next we evaluate the stability of the fixed points in the above
equations. As throughout our simulations $\overline{f}_D$ always
converges to $1/2$, we fix $\overline{f}_D =1/2$ and analyze the
consistency of the dynamical behavior of the equations with the
numerical results. It turns out that $m_D=(1-2q\overline{m}_D
)/2(1-q)$ is unstable for any value of $q$ and $\overline{m}_D$. As
for the stability of the other two equilibria, Fig. \ref{stab} shows
three regions in the (q,$\overline{m}_D$) plane, that we have named
$A$, $B$ and $C$ as they are in correspondence with the cases
described in the preceding section. In region $A$, both $m_D=0$ and
$m_D=1$ are stable, while in $B$ and $C$ only one of the two
equilibria are stable, respectively, $m_D=0$ and $m_D=1$. In regions
$B$ and $C$, moreover,  $m_D=(1-2q\overline{m}_D )/2(1-q)$ adopts
values outside the interval $[0,1]$. Consequently, for $q<1/2$ the
two extreme values of $m_D$ are stable. For $q>1/2$, on the other
hand, a departure of  $\overline{m}_D$ from $1/2$ can lead the
system to select only one of these two values. This analysis helps
us understanding why the state $A$ disappears for high values of
$q$, and  why the states $B$ and $C$ are not present when $q$ is
low.

As for the stability of the equilibria of Eq. (\ref{qf}), we find
that when $\overline{m}_D<2/3$ both $f_D=0$ and $f_D=1$ are stable
fixed points, while $f_D=1/2$ is unstable. When $\overline{m}_D>2/3$
the stability properties become inverted, and the only stable
equilibrium is $f_D=1/2$.

The compatibility of these analytical results with our previous
numerical results can be assessed as follows. In case $A$, we have
$\overline{m}_D=\overline{f}_D=1/2$, while the populations of both
males and females split into two groups adopting the extreme values
$0$ and $1$ for  $m_D$ and $f_D$. This agrees with the observed
results for low values of $q$. As $\overline{m}_D<2/3$, $f_D=0$ and
$f_D=1$ are stable. The results shown in Fig. \ref{stab}, in turn,
indicate that when  $q > 1/2$, fluctuations in the value of
$\overline{m}_D$ can make the system to depart from state $A$.

In case $B$, we have  $\overline{m}_D=0$ and  $\overline{f}_D=1/2$.
Accordingly,  the population of females splits in two groups
adopting the extreme values $0$ and $1$, while all males have
$m_D=0$. For $q>1/2$, fluctuations may lead the whole male
population to abandon the strategy $m_D=1$ and choose only $m_D=0$.
Again, as $\overline{m}_D<2/3$, $f_D=0$ and $f_D=1$ are stable.
However, when $q<1/2$ unless a specific initial condition is
selected, the system will converge to state $A$.

For case $C$ the analysis is  analogous to case $B$. Now, we have
$\overline{m}_D=m_D=1$ and  $\overline{f}_D=f_D=1/2$. When $q<1/2$,
$f_D=1/2$ is unstable, and $m_D$ splits between $0$ and $1$. Again,
when $q>1/2$ the system can converge to  $m_D=1$. Once $m_D>2/3$,
the only stable equilibrium is $f_D=1/2$. Despite the fact that this
analytical approach disregards ''microscopic´´ details of the
numerical simulations, thus, it correctly describes the collective
behavior of the two populations in the three cases, and for small
and large values of the imitation probability.

\section{Conclusion}
We have presented a simplified model to mimic the courtship process
in humans as a game. Inspired by empirical evidence, we have
proposed a coordination game where both players go better off by
choosing the same strategy, i.e. acting coordinately, but where at
the same time there are some constraints. Even when the best option
for both players is to act in a coordinated way, there  is no
coincidence as to which is the optimal individual strategy.
Consequently, coordination involves some kind of concession from at
least one of the players. This typical situation is represented in
the Battle of the Sexes. This game has been extensively analyzed due
to the interest rising from the fact that it has two Nash
equilibria, reflecting rational choices by the players. Rational
players would agree that the non-Nash outcomes are disadvantageous,
and that the Nash equilibria, even though one is inferior to the
other on each player's opinion, are the reasonable option. Neither
player would want to produce a disagreement that would lead to a
non-Nash outcome, even if this implies a conflict. In a population
of players, this rationality is well described by the replicator
dynamics.

In our case, instead, we were  interested at including non-rational
but still quantifiable motivations in our players. Thus, we proposed
a different dynamics, which still favors the possibility that the
players coordinate with each other in order to mate. After
unsuccessful interactions each player will modify the strategy
becoming indulgent toward the opposite sex, even at the cost of
moving off from the favorite strategy. In the absence of imitation,
the dynamics is myopic, as the collected information is only related
to the personal history.  The proposed dynamics turns out to be
non-payoff monotonic. This fact induces a dynamical behaviour
completely different from the one obtained from the replicator
equations. We verified the emergence of different dynamical
equilibria and the change on the stability of the previously
existing ones. Indeed, one of the most interesting features is that
the Nash equilibria are no longer stable under this new dynamics.
For some initial conditions, in fact, the dynamics leads the
populations towards the most unfavorable situation. Decisions
adopted by the players make them end up in an undesirable situation,
far from the ideal coordinated outcome.

Further inclusion of social influence, in the form of imitation
between males, adds to the behaviour of the system another
interesting feature. Tuning the intensity of the imitation tendency
we observe a transition between qualitatively different asymptotic
behavior. While the imitation tendency is low the populations of
both sexes split into similar groups that asymptotically approach
the extreme values. As the imitation intensity among males becomes
larger , the whole male population is led to a common pure strategy.
The choice of the final state is governed by microscopic
fluctuations in the initial distribution of strategies. Coordination
seems to be enhanced as the information is shared: while males adopt
a unified strategy, females fail to achieve the corresponding
coordinated strategy to maximize their collective  payoff. This
deviation from rationality is also to be ascribed to the non-payoff
monotonic character of the present courtship game.

\section{acknowledgement}
We acknowledge financial support from ANPCyT (PICT 2011-0545) and
from CONICET (PIP 11220080100076), Argentina.

\end{document}